\documentclass[mathleft]{an}
\usepackage{graphicx}
\usepackage{times}
\usepackage[colorlinks=true,citecolor=blue,linkcolor=blue]{hyperref}
\usepackage{amssymb,natbib}

\usepackage{fancyhdr}
\usepackage{longtable,lscape,amsmath}  
\usepackage{rotating}
\usepackage{xcolor}
\usepackage{natbib,twoopt}
\usepackage[T1]{fontenc}
\usepackage{aecompl}
\newcommand{\ltsima} {$\; \buildrel < \over \sim \;$}
\newcommand{\simlt} {\lower.5ex\hbox{\ltsima}} 
\newcommand{\gtsima} {$\; \buildrel > \over \sim \;$}
\newcommand{\simgt} {\lower.5ex\hbox{\gtsima}} 

\begin{document}

\Pagespan{789}{}
\Yearpublication{2015}%
\Yearsubmission{2015}%
\Month{?}%
\Volume{?}%
\Issue{?}%

\title{Radio Properties of the $\gamma$-ray Emitting CSO Candidate 2234+282}

\author{T. An\inst{1,2,4}\fnmsep\thanks{Corresponding author:
  \email{antao@shao.ac.cn}}
\and Y.-Z. Cui\inst{1,2}
\and K. \'{E}. Gab\'{a}nyi\inst{3,5} \thanks{speaker at the conference.}
\and S. Frey\inst{3}
\and W. A. Baan\inst{2}
\and W. Zhao\inst{2}
}
\titlerunning{$\gamma$-ray CSOs}
\authorrunning{An et al.}
\institute{ School of Electrical and Electronic Engineering, Shanghai Institute of Technology, 201418, Shanghai, China
\and
Shanghai Astronomical Observatory, Chinese Academy of Sciences, Nandan Road 80, Shanghai 200030, China
       \and
F\"{O}MI Satellite Geodetic Observatory, PO Box 585, H-1592 Budapest, Hungary
\and
Key Laboratory of Radio Astronomy, Chinese Academy of Sciences, Nanjing 210008, China
\and
Konkoly Observatory, MTA Research Centre for Astronomy and Earth Sciences, PO Box 67, 1525 Budapest, Hungary
}

\received{2015 Oct 19}
\accepted{2015 Oct 21}
\publonline{ }


\abstract{Most of the $\gamma$-ray emitting active galactic nuclei (AGN) are blazars, although there is still a small fraction of non-blazar AGN in the {\it Fermi}/LAT catalog. Among these misaligned $\gamma$-ray-emitting AGN, a few can be classified as Compact Symmetric Objects (CSOs).
In contrast to blazars in which $\gamma$-ray emission is generally thought to originate from highly beamed relativistic jets, the source of $\gamma$-ray emission in unbeamed CSOs remains an open question. The rarity of the $\gamma$-ray emitting CSOs is a mystery as well.
Here we present the radio properties of the $\gamma$-ray CSO candidate 2234+282.
}
\maketitle

\section{Introduction}

One of the intriguing discoveries of the {\it Fermi} Large Area Telescope (LAT)
survey is the overwhelming majority (98 \%) of the {\it Fermi}-detected AGN classified as blazars\\ \citep{Ackermann15}.
In leptonic models, $\gamma$-ray radiation from blazars is mostly explained by Compton scattering of low-energy synchrotron photons by the same relativistic electrons in the jets producing the synchrotron emission at lower frequencies \citep[e.g.,][]{BM96}; in hadronic models, the high-energy emission is dominated by relativistic protons \citep[e.g.,][]{Bottcher10}.
Besides the blazars, there are dozens of misaligned AGN detected by the {\it Fermi} \citep[e.g.,][]{Abdo10a,Ackermann15}. \\
The sample size is still small, though they constitute a very interesting class of $\gamma$-ray emitters. Cen A (NGC 5128), Per A (NGC 1275) and M87 (3C 274) are three well-studied nearby radio galaxies in which 
$\gamma$-ray emission was observed in the core. 
A single-zone synchrotron self-Compton (SSC) jet model was successfully applied to fit the LAT $\gamma$-ray spectrum of these radio galaxies, although other explanations exist \citep[e.g.][]{Abdo09}. 
These non-blazar gamma-ray AGN generally have moderate jet beaming, with Doppler boosting factor of 2 -- 4. 
Extended $\gamma$-ray emission was also detected from Cen A lobes \citep{Abdo10b},
and it is interpreted as the inverse Compton scattering of the cosmic microwave background photons by the relativistic electrons in the lobes.

Among the misaligned AGN, there is a sub-class named Compact Symmetric Objects (CSOs), which are regarded as young radio galaxies \citep[reviewed by][]{ODea98,Fanti09}. Theoretical models have predicted $\gamma$-ray emission from CSOs \citep{Stawarz08,Kino09,Migliori14}. 
However there are only two $\gamma$-ray CSO candidates known to date,
4C $+$55.17  \citep{McConville11} and PMN J1603$-$4904 \citep{Muller14}. 
In this paper, we report a new $\gamma$-ray CSO candidate 2234+282.

\section{Sample selection and data reduction}

In order to increase the sample size of $\gamma$-ray CSOs, we looked into the archival MOJAVE\footnote{http://www.physics.purdue.edu/astro/MOJAVE/index.html} \citep[Monitoring Of Jets in Active galactic nuclei with VLBA Experiments,][]{Lister09} database to search for new candidates.

The criteria of sample selection are : 
\begin{enumerate}
\item the source has been identified by {\it Fermi} with high statistical significance; 
\item the radio morphology is of CSO type, characterised with compact double lobes with or without a central core;
\item the total flux density which is integrated over the whole source in each VLBI image shows a steep or GHz inverted (GPS type) spectrum.
\end{enumerate}

In total, 344 AGN are cross-matched between the MOJAVE and {\it Fermi} catalogues.
Among them, we found only one source, 2234+282 matching the above three criteria.
Such a low percentage (0.3\%) is consistent with the rarity of the $\gamma$-ray CSOs in general.
Details of the radio emission properties are given in the next Section, including the spectral index $\alpha$ (defined as $S_\nu \propto \nu^\alpha$), the brightness temperature T$_\mathrm{b}$ of the core, and hot spot advance speed ($\beta_\mathrm{adv}$).

\begin{figure}
\centering
\includegraphics[width=0.4\textwidth]{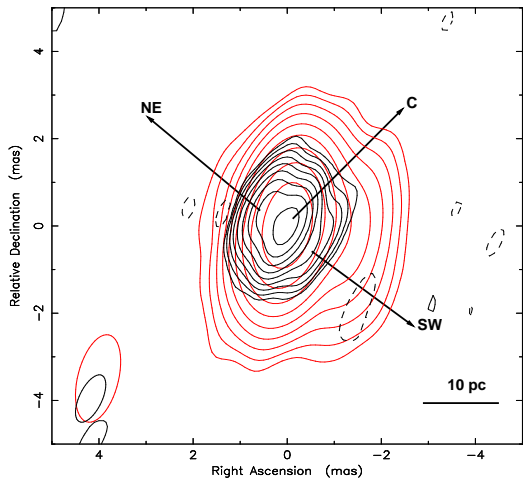}
\caption{Overlaid VLBI images of 2234$+$282 at 8.4 GHz (red contours) and 15 GHz (black contours). 
The 8.4-GHz image was made from the VLBA data on 1996 May 15 (project code: BB023), the restoring beam (shown as a red ellipse in the bottom left corner) is 2.05 mas $\times$ 0.87 mas at PA $= -12.8\degr$.   
The peak brightness is 0.708 Jy beam$^{-1}$.
The contour levels increase by a factor of 2, and the lowest contour is 1.5 mJy beam$^{-1}$ ($3\sigma$). 
The 15-GHz data were obtained on 1996 May 16 (project code: BK037).
The restoring beam is 1.17 mas $\times$ 0.49 mas at PA $= -23.4\degr$ (shown as a black ellipse in the bottom left corner).  
The peak brightness is 0.421 Jy beam$^{-1}$, and the lowest contour is 1.1 mJy beam$^{-1}$ ($3\sigma$). 
}
\label{fig1}
\end{figure}

\begin{figure}
\centering
\includegraphics[width=0.45\textwidth]{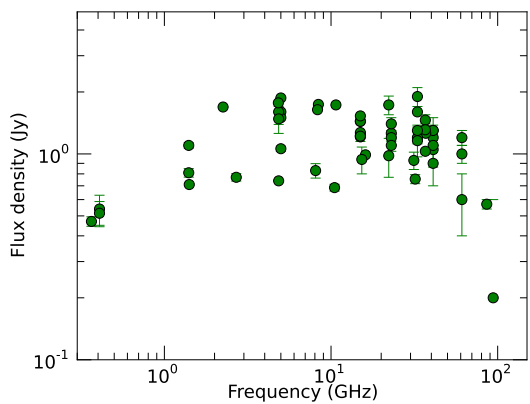}
\caption{Radio flux density versus frequency.
The data points, obtained from the single dish or VLA measurements, are collected from the NED.
The radio spectrum is complex above 1 GHz due to variability.
A significant decrease of flux density at 408 MHz suggests a low frequency turnover around 1 GHz.
}
\label{fig2}
\end{figure}

\section{CSO identification of 2234+282}

2234+282 (also named as CTD 135) has a redshift $z = 0.79$.
Its optical spectrum was described by \citet{Shaw12}  as transitional between flat-spectrum quasars and BL Lac type objects. 
It was detected by {\it Fermi} with a statistical significance of 52.1$\sigma$, corresponding to 3FGL J2236.3+2829 and 2FGL J2236.4+2828 \citep{Acero15}, but not detected by the Energetic Gamma Ray Experiment Telescope \citep[EGRET;][]{Hartman99} of the Compton Gamma Ray Observatory, probably due to insufficient sensitivity. 

The radio spectrum of the total flux density is complex above 1 GHz due to the variability of the non-simultaneous data points (Figure \ref{fig2})\footnote{http://ned.ipac.caltech.edu/}. Except for that, there is a significant decrease in the flux density at 408 MHz, suggesting a low frequency turnover around 1 GHz in the spectrum. 
The Very Large Array (VLA) observation performed at 1.5 GHz in 1984 revealed a bright compact component and a weak feature about 4.9\arcsec to the southwest \citep{Murphy93}.

\begin{figure*}
\centering
\includegraphics[width=1.0\textwidth]{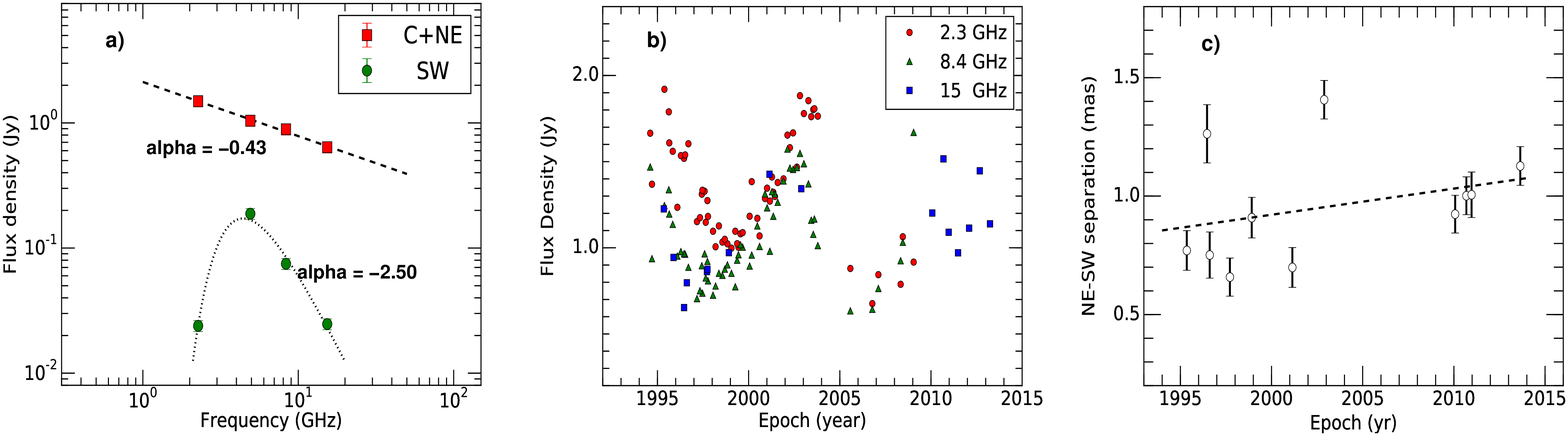}
\caption{Panel {\it a)} shows the radio spectra of the VLBI components in 2245+282.
The dashed line indicates the power law fit to the spectrum of the component {\tt C+NE}. 
The dotted line represents the model fit to the spectrum of {\tt SW} by using a power law together with the free-free absorption. We note that the spectrum can also be fitted with a synchrotron self-absorption model.
Panel {\it b)} displays the light curves at 2.3, 8.4 and 15 GHz. 
Panel {\it c)} shows the separation of components {\tt NE--SW} as a function of time. The dashed line denotes the least-squares fit to the observed data. 
}
\label{fig3}
\end{figure*}

The Very Long Baseline Array (VLBA) images at 2.3 GHz show a single compact component which is resolved along the northeast-southwest direction at frequencies higher than 5 GHz. 
Figure \ref{fig1} shows the 15-GHz image (black contours) overlaid on the 8.4-GHz one (red contours). 
At 15 GHz, the emission structure can be fitted with three Gaussian components, {\tt NE}, {\tt C} and {\tt SW}.  
Component {\tt C} is the most compact, and accounts for 65\% of the total flux density. 
We used circular Gaussians in the fitting to get the peak location and the size of the models, in order to simplify the comparison and to decrease the number of free parameters. 
One fifth of the beam size is taken as the error of the position.
Components {\tt NE} and {\tt C} are still blended at 8.4 GHz.
The combined {\tt NE+C} component shows a somewhat steep spectrum with a spectral index of $\alpha = -0.43 \pm 0.07$ (Figure \ref{fig3}-a).
The details of the VLBI observations used here to describe the spectral index of the source are given in Table \ref{t_VLBA}. The weaker {\tt SW} component shows an inverted spectrum with a turnover around 4 GHz. 
The spectrum of the optically thin section is rather steep, showing $\alpha = -2.50 \pm 0.18$. 

Since 2009 the MOJAVE program observations of the source were performed in full polarization. According to those the polarized emission is concentrated in components {\tt NE} and {\tt C}. 
Component {\tt C} does not show evident variation in polarization, but {\tt NE} shows significant variability in the polarized intensity at 15 GHz on a time scale of a few months. 

Components {\tt C} and {\tt NE} can be resolved only at 15 GHz and frequencies above. 
Since there has been no simultaneous observation at 15 and 24 GHz, and the source is variable (see below), an accurate estimation of the spectral index of component {\tt C} is not possible.
Even though, the central component {\tt C} is the brightest and most compact one among the three. 
The brightness temperature of {\tt C} ranges from $3.4 \times 10^{10}$ K to $1.5 \times 10^{12}$ K at 15 GHz. Thus {\tt C} is most likely the core.

Figure \ref{fig3}-b shows the light curves of the total flux density at 15, 8.4 and 2.3 GHz. 2.3 and 8.4 GHz observations were performed in the framework of the VLBA calibrator survey\footnote{the data are from the VLBI archive at http://astrogeo.org}. The flux densities at all three frequencies were calculated by summing up the CLEANed components. 
The source exhibits variability at all three frequencies with a percentage fluctuation index defined as 
\begin{equation}
u = 100 \frac{\sigma}{\langle S\rangle}
\end{equation}
\citep{Hee87}, in which $\langle S\rangle$ is the mean and $\sigma$ is the standard deviation.
That gives $u = 25.1\%$ (15 GHz), 25.9\% (8.4 GHz)  and 30.6\% (2.3 GHz). 
Since the central component {\tt C} contributes the majority of the total flux density, the variability may in fact represent the flux density variation in {\tt C}.

We investigated the position change of component {\tt NE} with respect to the assumed core {\tt C}, and found that {\tt NE} apparently moves toward the core {\tt C}. This is actually indicative that the component {\tt C} is a mixture of the core and an inner fast-moving jet (not resolvable at 15 GHz) which was ejected toward the northeast. 
We then used {\tt SW} component as the reference to calculate the hot spot separation speed. 
The angular separation between components {\tt NE} and {\tt SW} as a function of time is depicted in Figure \ref{fig3}-c.
The separation rate is estimated as $\beta_\textrm{sep} = 0.014  \pm 0.004 $ mas yr$^{-1}$, corresponding to a projected hot spot separation speed of $0.6 \pm 0.2 \,c$.
Assuming that the jet and counterjet have an equal speed, the average hot spot advance speed is $\sim 0.3\,c$. 
This hot spot speed is a typical value for CSOs \citep{AB12}, but much lower than the jet speed in blazars.

To summarize, all pieces of evidences, including the \\ compact triple structure, steep spectrum, and low jet speed are consistent with each other to imply that 2234+282 is a CSO. The brightness temperature and moderate variability of the core indicate that the innermost jet is Doppler boosted.

\renewcommand{\tabcolsep}{0.13cm}
\begin{table}
\caption{Interferometric observations used to describe the spectral index of the source (Figure \ref{fig3} Panel {\it a)}).}
\label{t_VLBA}
\begin{tabular}{cccccc}\hline
Epoch & $\nu$ & Project id & S$_{C+NE}$ & S$_{SW}$ \\ 
 & (GHz) & & (Jy) & (Jy) \\
\hline
1996 May 15 & 2.3 & BB023 & $1.49\pm0.15$ & $0.02\pm0.002$\\
1996 Jun 5  & 5   & BH019 & $1.04\pm0.10$ & $0.18\pm0.02$ \\
1996 May 15 & 8.4 & BB023 & $0.89\pm0.09$ & $0.07\pm0.01$ \\
1996 May 16 & 15  & BK037 & $0.64\pm0.06$ & $0.02\pm0.002$\\
\hline
\end{tabular}
\end{table}

\section{Discussion}

The location and the radiation mechanism of $\gamma$-ray emission from CSOs remain open questions.
A single-zone SSC emission model well fits the broadband SEDs from radio up to GeV energy bands in blazars and also in some misaligned AGN \citep[e.g., M87:][]{Abdo09}. 
The one-zone SSC model requires relativistic beaming. 
However, unlike the blazar jets, the CSO jets are in general mildly relativistic and unbeamed \citep{AB12}. 
It is obvious that this SSC model can not reproduce the $\gamma$-ray emission from CSOs. 
Some other models have been investigated to interpret the $\gamma$-ray radiation from CSOs.
For example,
\citet{Stawarz08} proposed that the relativistic electrons injected from the hot spots to the CSO lobes can up-scatter the ultraviolet photons from the disk to GeV energy range. 
\citet{Migliori14} proposed a two-zone SSC jet model for  the $\gamma$-ray emission from GPS and CSS quasars, in which the seed synchrotron photons which are from an inner blazar-like jet knot are up-scattered by the electrons in an outer slower jet knot.
In addition, bremsstrahlung emission from the shocked plasma in the cocoon of CSO  may also have a significant contribution to the high-energy emission \citep{Kino09}.
The model predicts $\gamma$-ray emission from CSOs that could be detected by {\it Fermi}, since the electron density and temperature are much higher in the young radio lobes than those in older ones.
The difference of these models is that the first two \citep{Stawarz08,Migliori14} are of non-thermal origin and and the last one \citep{Kino09} is thermal. 

The observational properties of 2234+282 make it a $\gamma$-ray CSO candidate, increasing the number of $\gamma$-ray CSO candidates from 2 to 3. 
The number of CSOs detected in GeV energy range so far is significantly lower than predicted by theoretical models. \cite{Filippo15} \\ searched 60 CSOs and analyzed 6 years of LAT data but failed to confirm any $\gamma$-ray CSO.
The theoretical calculations show that the model-predicted $\gamma$-ray flux is near or below the current {\it Fermi} detection limit \citep{Migliori14}.
Continuing search of a complete CSO sample based on the {\it Fermi} data at a lower energy range of 0.1 -- 10 GeV is underway \citep{Filippo15} to verify whether the CSO emission is intrinsically dominated in low energy band.
On the other hand, the morphology, spectral properties and kinematics of these three sources suggest mildly relativistic jet. 
VLBI data of $\gamma$-ray CSO candidates provide important information {\bf on} the radio jet luminosity and jet speed, placing stringent constraints on the $\gamma$-ray emission models. 
Complementary wide-band SED fits of these three sources are crucial for verifying whether the jet plays an important role in generating $\gamma$-ray emission from CSOs \citep[the model of \\][]{Migliori14}.
If however the $\gamma$-ray emission from CSOs is found to be of thermal origin and related to the lobes, that would identify a new $\gamma$-ray emission mechanism for AGN as proposed by \citet{Kino09}.

\acknowledgements
The research is supported by the European Commission Seventh Framework Programme (FP/2007-2013) under grant agreement No 283393 (RadioNet3).
This research is supported by the Chinese Ministry of Science and Technology grant (2013CB837900), Shanghai Rising-Star Program, the Hungarian Scientific Research Fund (OTKA NN110333) and the\,China--Hun\-gary Collaboration and Exchange Programme by the International Cooperation Bureau of the Chinese Academy of Sciences.
This research has made use of data from the MOJAVE database
that is maintained by the MOJAVE team.
The NRAO is a facility of the National Science Foundation operated under cooperative agreement by Associated Universities, Inc.


\begin{thebibliography}{}
\bibitem[\protect\citeauthoryear{Abdo et al.}{2009}]{Abdo09} Abdo, A.A., Ackermann, M., Ajello, M., et al. 2009, ApJ, 707, 55 
\bibitem[\protect\citeauthoryear{Abdo et al.}{2010a}]{Abdo10a} Abdo, A.A., Ackermann, M., Ajello, M., et al. 2010a, ApJ, 720, 912 
\bibitem[\protect\citeauthoryear{Abdo et al.}{2010b}]{Abdo10b} Abdo, A.A., Ackermann, M., Ajello, M., et al. 2010b, Science, 328, 725
\bibitem[\protect\citeauthoryear{Acero et al.}{2015}]{Acero15} Acero, F., Ackermann, M., Ajello, M., et al. 2015, ApJS, 218, 23  
\bibitem[\protect\citeauthoryear{Ackermann et al.}{2015}]{Ackermann15} Ackermann, M., Ajello, M., Atwood, W.\,B., et al. 2015, ApJ, 810, 14 
\bibitem[\protect\citeauthoryear{An \& Baan}{2012}]{AB12} An, T., \& Baan W.A. 2012, ApJ, 760, 77
\bibitem[\protect\citeauthoryear{Bloom \& Marscher}{1996}]{BM96} Bloom \& Marscher 1996, ApJ, 461, 657
\bibitem[\protect\citeauthoryear{B\"{o}ttcher}{2010}]{Bottcher10} B\"{o}ttcher, M. 2010, in Proc. Fermi Meets Jansky, MPIfR, Bonn, eds. T. Savolainen, E. Ros, R. W. Porcas, \& J. A. Zensus, p. 41
\bibitem[\protect\citeauthoryear{D'Ammando et al.}{2015}]{Filippo15} D'Ammando, F., Orienti, M., Giroletti, M., et al. 2015, these proceedings
\bibitem[\protect\citeauthoryear{Fanti}{2009}]{Fanti09} Fanti, C., 2009, AN, 330, 120
\bibitem[\protect\citeauthoryear{Hartman et al.}{1999}]{Hartman99} Hartman, R.C., Bertsch, D.\,L., Bloom, S.\,D., et al. 1999, ApJS, 123, 79
\bibitem[Heeschen et al.(1987)]{Hee87} Heeschen, D.~S., 
Krichbaum, T., Schalinski, C.~J., \& Witzel, A.\ 1987, \aj, 94, 1493 
\bibitem[\protect\citeauthoryear{Kino et al.}{2009}]{Kino09} Kino M., Ito, H., Kawakatu, N., \& Nagai, H. 2009, MNRAS, 395, L43
\bibitem[\protect\citeauthoryear{Lister et al.}{2009}]{Lister09} Lister M., Homan, D.\,C., Kadler, M., et al. 2009, ApJ, 696, L22
\bibitem[\protect\citeauthoryear{McConville et al.}{2011}]{McConville11} McConville, W., Ostorero, L., Moderski, R., et al. 2011, ApJ, 738, 148
\bibitem[\protect\citeauthoryear{Migliori et al.}{2014}]{Migliori14} Migliori G., Siemiginowska, A., Kelly, B.\,C., et al. 2014, ApJ, 780, 165
\bibitem[\protect\citeauthoryear{M\"{u}ller et al.}{2014}]{Muller14} M\"{u}ller C., Kadler, M., Ojha, R., et al. 2014, A\&A, 559, 115
\bibitem[\protect\citeauthoryear{Murphy et al.}{1993}]{Murphy93} Murphy D.W., Browne I.W.A., Perley R.A. 1993, MNRAS, 264, 298
\bibitem[\protect\citeauthoryear{O'Dea}{1998}]{ODea98} O'Dea C.P. 1998, PASP, 110, 493
\bibitem[\protect\citeauthoryear{Shaw et al.}{2012}]{Shaw12} Shaw M.S., Romani, R.\,W., Cotter, G., et al. 2012, ApJ, 748, 49
\bibitem[\protect\citeauthoryear{Stawarz et al.}{2008}]{Stawarz08} Stawarz, \L., Ostorero, L., Begelman, M.\,C., et al. 2008, ApJ, 680, 911

\end{thebibliography}
\end{document}